# MALDI-TOF and Quantum Chemical Study of Non-stoichiometric Tantalum Oxychloride Clusters

E. G. Il'in[1], A.S. Parshakov[1], V.G. Yarzhemsky[1,2], E.A. Teplyakov[1,2,3], A.K. Buryak[4]

1) Kurnakov Institute of General and Inorganic Chemistry of RAS, Leninsky pr. 31, Moscow, 119991 Russia
2) Moscow Institute of Physics and Technology, Dolgoprudny, 9 Institutsky lane, 141700, Moscow Region, Russia
3) Institute for High Pressure Physics of RAS, 142190 Troitsk, Russia
4) Frumkin Institute of Physical Chemistry and Electrochemistry of RAS, Leninsky pr. 31, Moscow, 119991 Russia

**Summary**

Using the MALDI-TOF spectroscopy method, clusters of non-stoichiometric tantalum oxychlorides formed as a result of hydrolysis of nanoscale reaction products of tantalum pentachloride with acetylene and of interaction with oxygen and air moisture in the process of sample preparation were studied. The formation of oxychloride clusters containing from 2 to 15 tantalum atoms is shown. Quantum chemical calculations of the structure and relative thermodynamic stability of the possible isomers are carried out. The stability of cage Ta structures connected by oxygen bridges with significant contribution of metal-metal bond is established

**Introduction**

The combination of the experimental MALDI-TOF mass spectrometry with quantum chemical calculations allows us not only to identify the composition of multinuclear transition metal clusters in the gas phase, but also to draw conclusions about the structure and relative thermodynamic stability of their possible geometric isomers. Previously, we used this approach to study the catalyst for linear polymerization of acetylene and substituted alkynes based on molybdenum halides. It was shown that the active centers of the catalyst are 13-atom organometallic clusters of lower molybdenum chlorides[1-3]. Lower Mo oxides, in particular $MoO_2$, are used as functional materials in catalysis, in sensors, solar cells[4-7]. A combination of these methods was used to study $MoO_2$ containing a nanosized fraction obtained by thermal decomposition of the complex of molybdenum ion with N-substituted hydroxylamine $[MoO_2 (i-C_3H_7NHO)_2]$[8]. The composition in the gas phase of cation clusters of lower non-stoichiometric Mo oxides was established and stability of symmetric cell structures was shown[9]. Clusters of anions of higher Mo oxides were also studied[10].

Niobium and tantalum pentachlorides are effective catalysts for the trimerization of acetylene and the stereoselective cyclotrimerization of monosubstituted alkynes $HC≡CR$[11-14], however, information on the study of catalytically active complexes is mainly limited by their composition[15]. We assumed that the niobium catalyst is also an organometallic cluster of lower niobium chlorides, however, when studying catalysts with the composition NbCl$_2$ ± 0.1 (Cn ± 1Hn ± 1) (n = 10-12) by MALDI TOF mass spectrometry in the mode of identification of negative ions. In the gas phase, previously unknown clusters of niobium oxychlorides were identified[16]. The latter were formed as a result of the active interaction of the cyclotrimerization catalyst of alkynes NbCl$_{2\pm0,1}$(C$_{n\pm1}$H$_{n\pm1}$) (n=10÷12) with oxygen and air moisture during sample preparation due to its high dispersion.

On the other hand, the investigations of tantalum clusters are rather scant. Small metal oxide clusters of V, Nb and Ta were investigated by mass-spectrometry and infra-red spectroscopy[17]. Systematic quantum chemical study of tetra-nuclear metal oxide clusters $M_4O_n^{-/0}$ (M=Nb,Ta) resulted $T_d$ structure for stoichiometric $M_4O_{10}$ clusters[18]. Transition-metal oxide clusters of the $M_nO_m^+$ (M=V,Nb,Ta) were produced by laser vaporization in a pulsed nozzle cluster source and time-of-flight mass spectrometry detected that the most stable are $M_2O_4^+$, $M_3O_7^+$, $M_4O_9^+$, $M_5O_{12}^+$, $M_6O_{14}^+$ and $M_7O_{17}^+$ [19]. These clusters have stable bonding networks at their core and additional excess oxygen at their periphery. Investigations of formation and distribution of neutral vanadium, niobium and tantalum oxide clusters by mass-spectrometry and single photon ionization revealed that for tantalum and niobium families with even number of metal atoms, oxygen deficient clusters have the general formula $(MO_2)_2(M_2O_5)_y$ [20]. In According to X-ray studies octahedral Ta$_6$ clusters with twelve Cl atoms in edge positions and six Cl atoms in apical positions form the parts of Cs$_2$PbTa$_6$Cl$_{18}$ and CsPbTa$_6$Cl$_{18}$ compounds[21].

The aim of present work was to study the tantalum oxychloride clusters making use of the combination of MALDI-TOF spectroscopy and quantum chemical calculations. Considering the similarity of the properties of elements - analogues of niobium and tantalum, as an initial material, we synthesized a tantalum catalyst, obtained similarly by the reaction of tantalum pentachloride with acetylene. We believed that due to its high dispersion, the latter will also actively interact during sample preparation with oxygen and air moisture with the formation of tantalum oxychlorides.

**Experimental**

For the synthesis of the catalyst, acetylene was passed through a saturated solution of TaCl$_5$ in benzene (0.0174 mol /L) until the evolution of HCl ceased. The dark blue precipitate was washed with solvent and dried under vacuum. According to chemical analysis, the substance contains 34.9% Ta, 25.5% Cl, 34.7% C, 4.3% H, which corresponds to the composition of $TaCl_{3.7}C_{15}H_{22}$, which is close to theoretical composition $TaCl_4C_{16}H_{16}$.

MALDY TOF mass spectra were recorded on a Bruker Ultraflex instrument equipped with a nitrogen laser with a wavelength of 337 nm and energy of 110 μJ., at the Center of Collective Use of Frumkin Institute of Physical Chemistry and Electrochemistry of RAS. The mass range is 20 - 4500 Da. Mass spectra were recorded at a laser energy of 80 - 95% of the maximum. Laser "shots" were carried out in 25 different sections of the test sample. A double-sided tape was placed on the surface of the stainless steel target, on which the powder of the test sample was applied. Based on the molecular mass picture of the isotopic distribution, the formulas of the intended compounds corresponding to the m/z ratios were calculated. The MALDI TOF mass spectrum recorded in the mode of detection of negative ions contained anions of tantalum oxychlorides (Fig. 1).

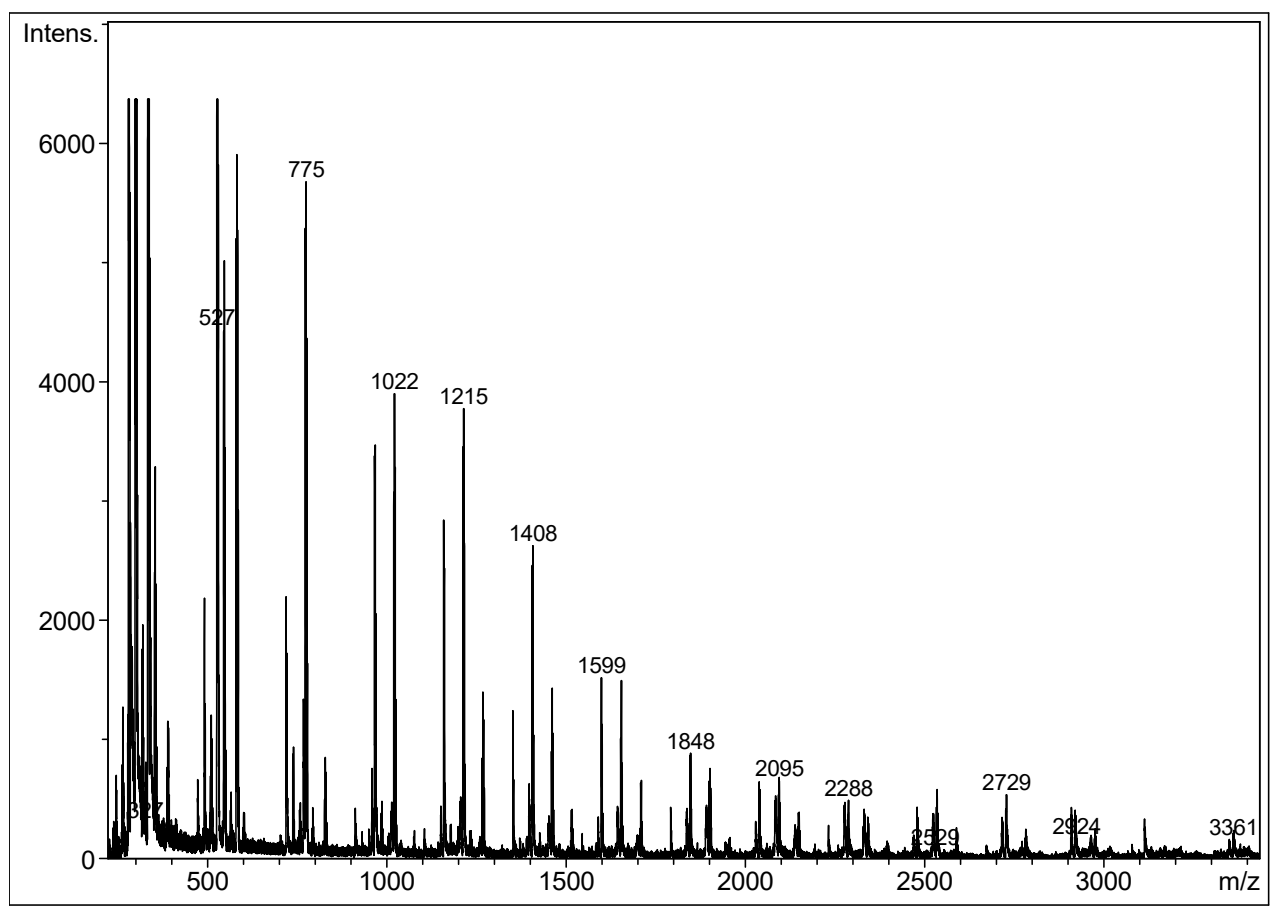

Figure 1. MALDI-TOF spectrum of HTaOCl clusters

Table 1. Experimental *m/z* values and corresponding chemical compositions of clusters

| compositions | m/z | compositions | m/z |
|---|---|---|---|
| $H_2Ta_3O_7Cl_2^-$ | 727 | $H_2Ta_6O_{15}Cl^-$ | 1364 |
| $H_2Ta_3O_6Cl_4^-$ | 781 | $H_2Ta_6O_{14}Cl_3^-$ | 1418 |
| $H_2Ta_4O_{10}Cl^-$ | 921 | $H_2Ta_6O_{13}Cl_5^-$ | 1472 |
| $H_2Ta_4O_9Cl_3^-$ | 975 | $H_3Ta_7O_{18}^-$ | 1558 |
| $H_2Ta_4O_8Cl_5^-$ | 1029 | $H_3Ta_7O_{17}Cl_2^-$ | 1612 |
| $H_3Ta_5O_{10}^-$ | 1116 | $H_3Ta_7O_{16}Cl_4^-$ | 1666 |
| $H_2Ta_5O_{12}Cl_2^-$ | 1169 | $H_3Ta_7O_{15}Cl_6^-$ | 1722 |
| $H_2Ta_5O_{11}Cl_4^-$ | 1225 | $H_4Ta_8O_{18}Cl_5^-$ | 1914 |

**Calculation method**

Calculations were performed by making use of GAMESS package[22] with the Becke's three-parameter hybrid exchange[23] and the Lee-Yang-Parr gradient-corrected correlation functional (B3LYP)[24] and basis set LanL2DZ for Ta, O and H and G-31(d) for Cl. Only stable structures i.e. without imaginary frequencies are discussed throughout.

**Calculation results**

Stable structures of $H_2Ta_3O_6Cl_4^-$ and relative total energies obtained by quantum chemical calculations are presented in Figures 2 a), b), c) and d) and theoretical bond lengths are presented in Table 2. The most stable structures are presented first throughout. It is seen from Figure 2 that cyclic structures are more stable that linear ones. Among cyclic structures the more stable is one with O-atom over a plane connected by almost equal bonds (2.13 Å) with three Ta atoms (Fig.2 a). The structure with Cl over a plane connected with two shorter bond (2.70 Å) and one longer bond (3.70 Å) with Ta atoms is less stable. Amongst the linear structures the structure with single oxygen bridges is more stable than that with double oxygen bridges The Ta-Ta bonds lengths in cyclic structure Fig 2 a) (2.94 $\pm$ 0.11 Å) are significantly shorter than in structure Fig 2 b) (3.516 $\pm$ 0.06 Å). The most stable cyclic structure of $H_2Ta_3O_7Cl_2^-$ are shown in Figure 3. Figures 4 a) and 4 b) show molecular orbitals corresponding to the three centre metal bonds for cyclic $Ta_3$ structures. Thus the most energetically stable $Ta_3$ clusters reveal three center metal-metal bond.

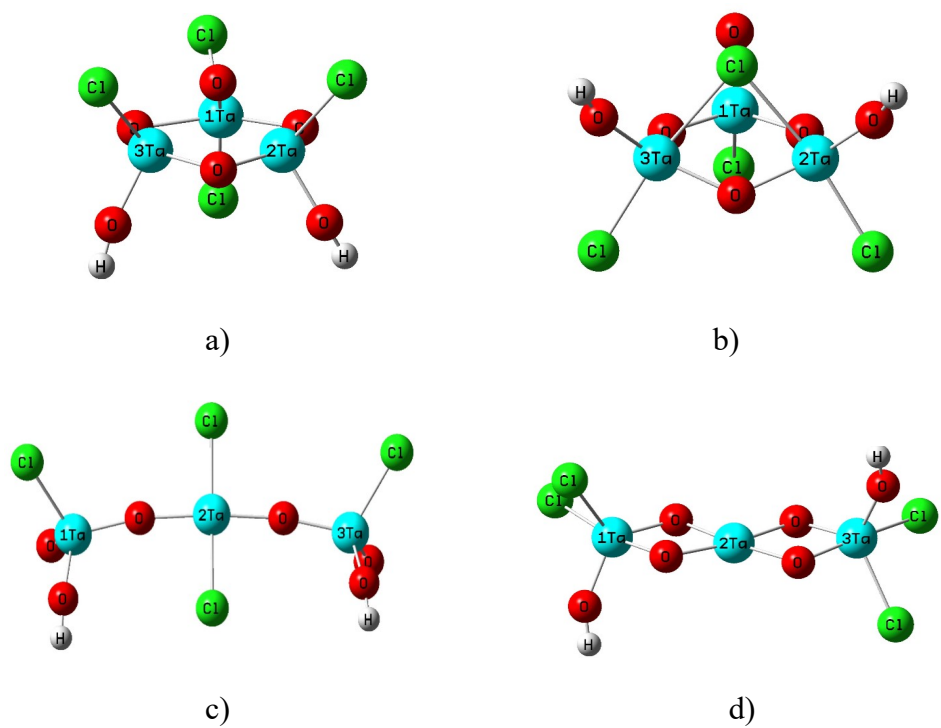

a)   b)

c)   d)

Figure 2. Theoretical stable structures of $H_2Ta_3O_6Cl_4^-$ and their relative total energies (in parenthesis) a)(0.00 eV), b)(+0.53eV), c)(+0.62eV) and d)(+0.95eV).

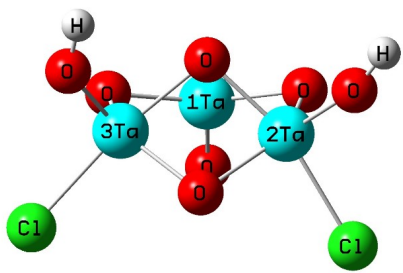

Figure 3. Stable structure of $H_2Ta_3O_7Cl_2^-$

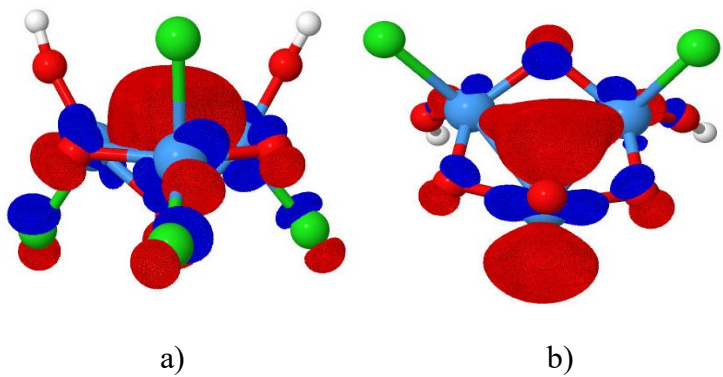

a)   b)

Figure 4. MOs with metal-metal bonds of structures a) $H_2Ta_3O_6Cl_4^-$ (Fig. 2 a) and b) $H_2Ta_3O_7Cl_2^-$ (Fig. 3)

Table 2. Interatomic distances d(Å) in Ta$_3$ clusters.

| Figure | Ta-Ta | Ta-Cl | Ta-O$_{bridge}$ | Ta-OH |
|---|---|---|---|---|
| Fig 2 a) | 2.9(1) | 2.382(3) | 1.94(7) | 1.877(3) |
| Fig 2 b) | 3.52(6) | 2.388(9) | 1.925(4) | 1.84(6) |
| Fig 2 c) | 3.811(1) | 2.36(2) | 1.91(5) | 1.82(7) |
| Fig 2 d) | 3.034 | 2.39(2) | 1.96(7) | 1.870(1) |
| Fig 3 | 3.02(5) | 2.394(6) | 1.97(3) | 1.85(7) |

In the case of Ta$_4$ clusters mass-spectra resulted in four clusters in the same metal oxidation state, namely $H_2Ta_4O_{10}Cl^-$, $H_2Ta_4O_9Cl_3^-$ and $H_2Ta_4O_8Cl_5^-$. In this case we obtained two possible structures: distorted tetrahedron and quadrangle. In all cases tetrahedral structures are more stable than quadrangle. The stable structures are shown in Figures 5-7 and bond length are presented in Table 3 and Table 4. The most stable in all cases are distorted tetrahedral structures. Figure 8 a) and 8 b) show MO with a pronounced metal-metal bond for tetrahedral $H_2Ta_4O_8Cl_5^-$ and $H_2Ta_4O_9Cl_3^-$ respectively. The metal-metal distance in less stable quadrangle structure and there are no orbitals with three centre meal-metal bonds.

Table 3. Interatomic distances d(Å) in tetrahedral structures of Ta$_4$.

| Figure | Ta$_{eq}$-Ta$_{eq}$ | Ta$_{eq}$-Ta$_{ax}$ | Ta-Cl | Ta-O$_{bridge}$ | Ta-OH |
|---|---|---|---|---|---|
| Fig 5 a) | 3.08(7) | 3.52(7) | 2.40(3) | 1.95(7) | 1.880(8) |
| Fig 6 a) | 3.11(9) | 3.38(9) | 2.42(5) | 1.95(4) | 1.84(5) |
| Fig 7 a) | 3.2(2) | 3.49(5) | 2.449 | 1.95(4) | 1.83(6) |

Table 4. Interatomic distances d(Å) in quadrangle structures of Ta$_4$ clusters.

| Figure | Ta-Ta | Ta-Cl | Ta-O$_{bridge}$ | Ta-OH |
|---|---|---|---|---|
| Fig 5 b) | 3.547(1) | 2.38(3) | 1.93(4) | 1.81(5) |
| Fig 6 b) | 3.58(4) | 2.39(3) | 1.97(2) | 1.85(6) |
| Fig 7 b) | 3.4(2) | 2.386 | 1.94(3) | 1.82(7) |

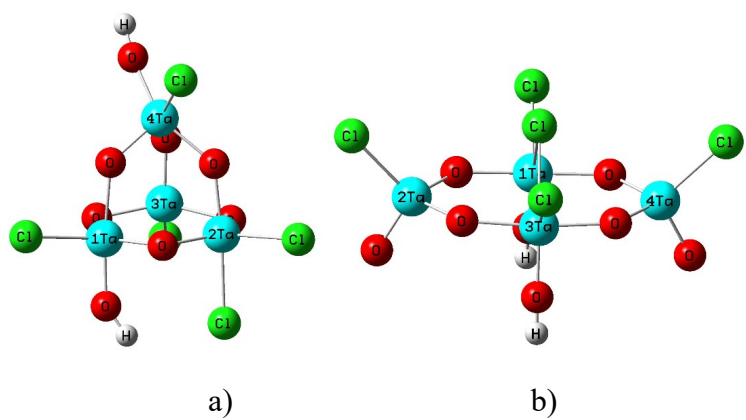

Figure 5. Stable structures of $H_2Ta_4O_8Cl_5^-$ and their relative total energies (in parenthesis) a) distorted tetrahedron (0.00 eV) and b) quadrangle (+0.92eV).

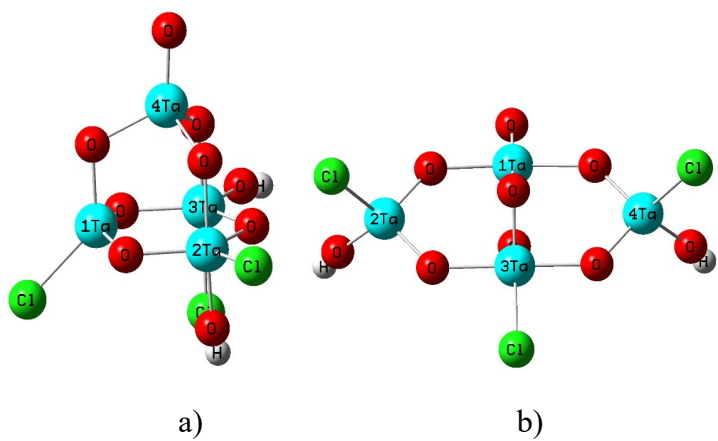

Figure 6. Stable structures of $H_2Ta_4O_9Cl_3^-$ and their relative total energies (in parenthesis) a) distorted tetrahedron (0.00 eV) b) quadrangle (+1.97eV).

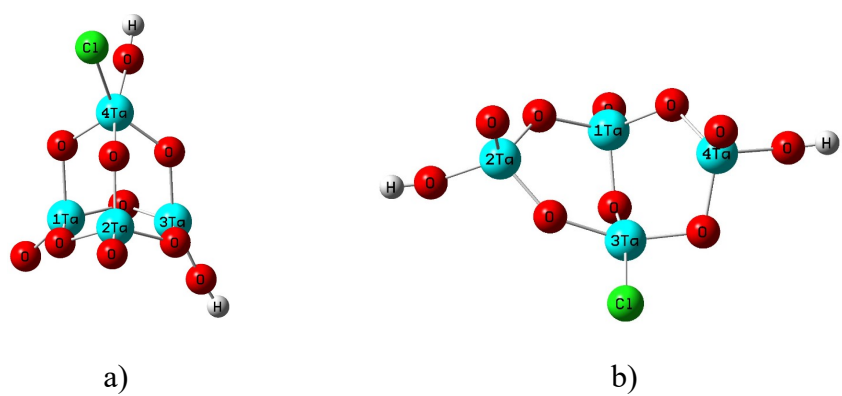

Figure 7. Stable structures of $H_2Ta_4O_{10}Cl^-$ and their relative total energies (in parenthesis) a) distorted tetrahedron (0.00 eV) b) distorted quadrangle (+0.12eV).

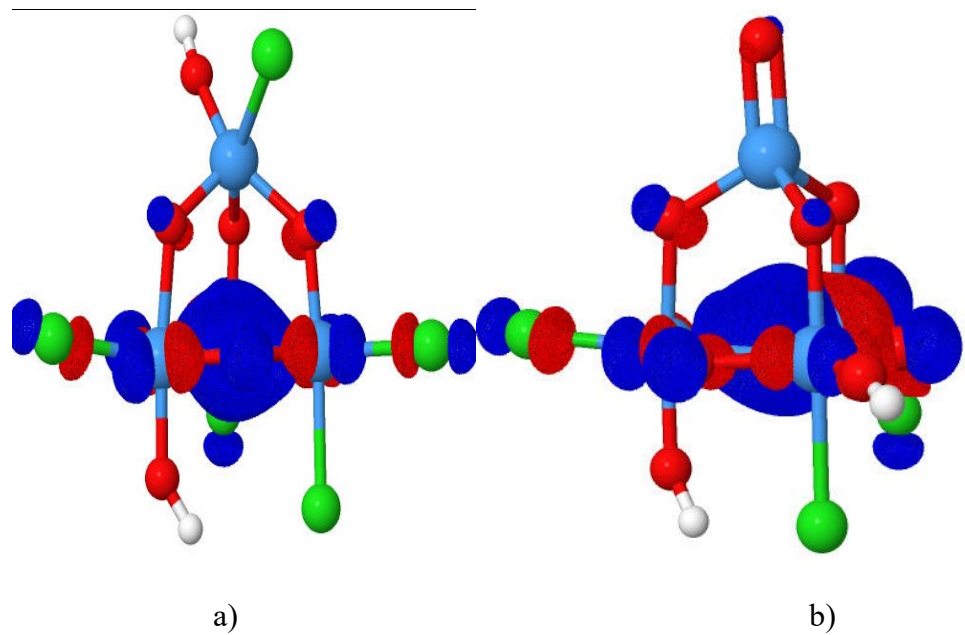

a)                                                      b)

Figure 8. Orbitals representing metal-metal bonds in tetrahedral Ta$_4$ structures a) $H_2Ta_4O_8Cl_5^-$ b) $H_2Ta_4O_9Cl_3^-$

There are three Ta$_5$ cluster in MALDI spectra, namely $H_3Ta_5O_{10}^-$, $H_2Ta_5O_{11}Cl_4^-$ and $H_2Ta_5O_{12}Cl_2^-$ The oxidation states of Ta in the last two clusters are larger than in the first one. In all cases we investigated two initial forms – trigonal bipyramid and tetragonal pyramid. Stable tetragonal pyramid structure was obtained for $H_2Ta_5O_{11}Cl_4^-$ only, but its total energy is larger then that for trigonal bipyramid. The structures are presented in Figures 9-11 and examples of metal-metal bonds in Figure 12. Bond lengths are presented in Table 5. Figure 12 shows metal-metal bond, which include four Ta atoms (Fig. 12 a), one equatorial and one polar Ta atoms (Fig.12 b) three equatorial Ta atom (Fig.12 c).

Table 5. Interatomic distances d(Å) in Ta$_5$ clusters.

| Figure | Ta$_{eq}$-Ta$_{eq}$ | Ta$_{eq}$-Ta$_{ax}$ | Ta-Cl | Ta-O$_{bridge}$ | Ta-OH |
|---|---|---|---|---|---|
| Fig 9 | 2.83(6) | 3.07(2) |  | 1.96(1) | 1.94(2) |
| Fig 10 a) | 3.106(2) | 3.44(3) | 2.402(4) | 1.95(2) | 1.886(1) |
| Fig 10 b) | 3.41(3) | 3.38(5) | 2.37(2) | 1.940(7) | 1.892(7) |
| Fig 11 | 3.127(5) | 3.387(1) | 2.406(1) | 1.95(4) | 1.891(1) |

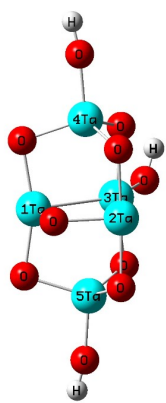

Figure 9. Bipyramid structure of $H_3Ta_5O_{10}^-$.

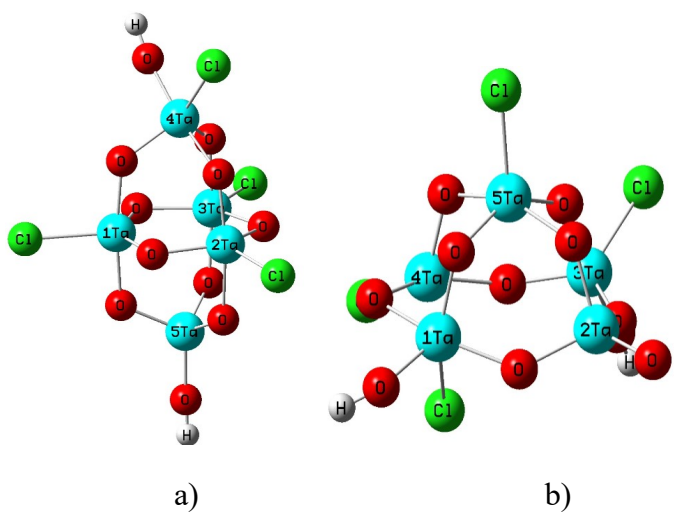

a)                  b)

Figure 10. Stable structures of $H_2Ta_5O_{11}Cl_4^-$ and their relative total energies (in parenthesis) a) bipyramid structures (0.00 eV) and b) tetragonal pyramid (+0.96eV).

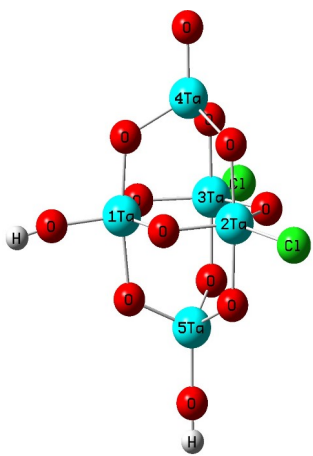

Figure 11. Bipyramid structure of $H_2Ta_5O_{12}Cl_2^-$.

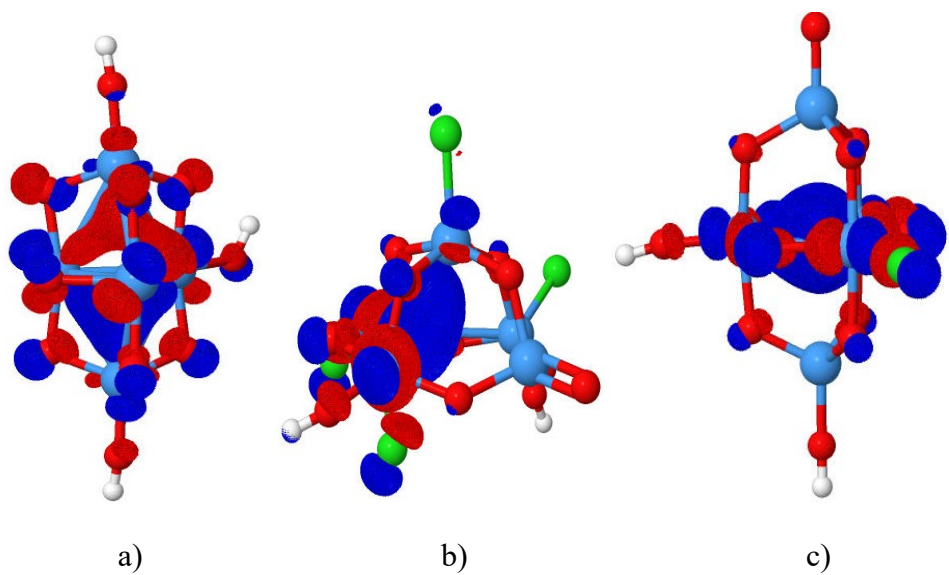

| | | |
|---|---|---|
| a) | b) | c) |

Figure 12. Orbitals representing metal-metal bonds in some Ta$_5$ structures. a) $H_3Ta_5O_{10}^-$ b) $H_2Ta_5O_{11}Cl_4^-$ c) $H_2Ta_5O_{12}Cl_2^-$

Table 6. Interatomic distances d(Å) in octahedrons structures of Ta$_6$ clusters.

| Figure | Ta$_{eq}$-Ta$_{eq}$ | Ta$_{eq}$-Ta$_{ax}$ | Ta-Cl | Ta-O$_{bridge}$ | Ta-OH |
|---|---|---|---|---|---|
| Fig 12 a) | 3.479(3) | 3.476(1) | 2.325(1) | 1.945(1) | 1.862(1) |
| Fig 13 a) | 3.41(6) | 3.44(2) | 2.331(1) | 1.949(5) | 1.865(1) |
| Fig 14 b) | 3.403(1) | 3.48(2) | 2.335(1) | 1.955(1) | 1.870 |

Table 7. Interatomic distances d(Å) in bipyramids structures of Ta$_6$ clusters.

| Figure | Ta$_{eq}$-Ta$_{eq}$ | Ta$_{eq}$-Ta$_{ax}$ | Ta-Cl | Ta-O$_{bridge}$ | Ta-OH |
|---|---|---|---|---|---|
| Fig 12 b) | 3.454(5) | 3.612(6) | 2.37(2) | 1.948(8) | 1.861(1) |
| Fig 13 b) | 3.48(4) | 3.56(4) | 2.348(7) | 1.950(7) | 1.864(1) |
| Fig 14 a) | 3.48(1) | 3.56(4) | 2.334 | 1.955(4) | 1.871(1) |

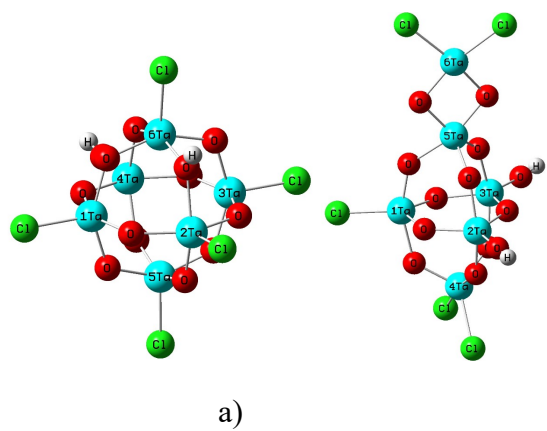

Figure 13. Stable structures of $H_2Ta_6O_{13}Cl_5^-$ and their relative total energies (in parenthesis) a)octahedron structure(0.00 eV) b) bipyramid structure (+2.36 eV).

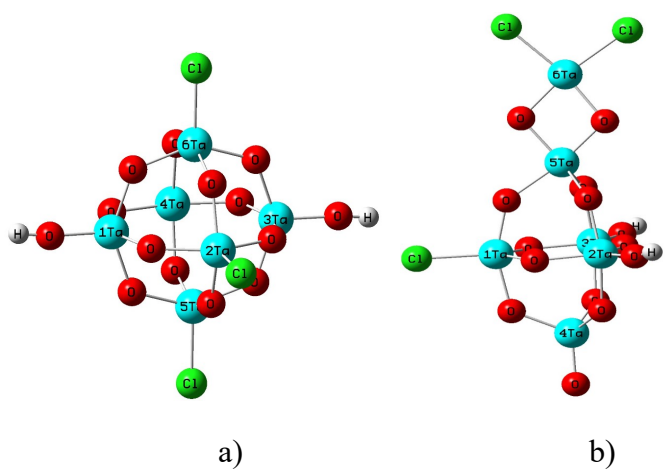

Figure 14. Stable structures of $H_2Ta_6O_{14}Cl_3^-$ and their relative total energies (in parenthesis) a)octahedron structure (0.00 eV) b) bipyramid structure (+1.87eV).

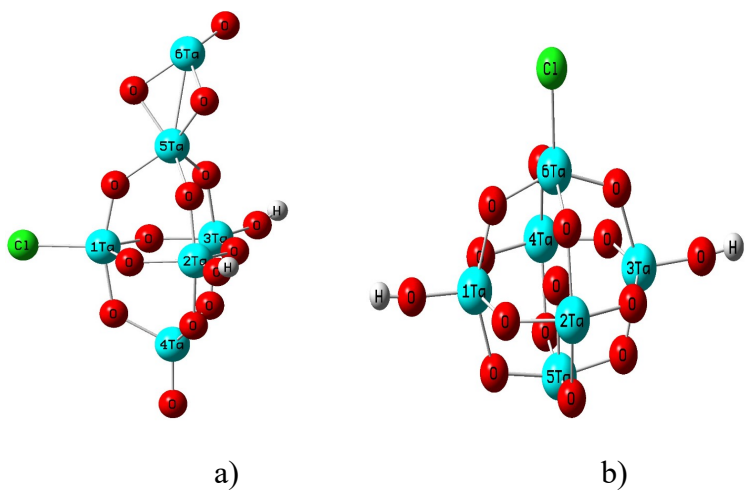

Figure 15. Stable structures of $H_2Ta_6O_{15}Cl^-$ and their relative total energies (in parenthesis) a)bipyramid structure (0.00 eV) b) octahedron structure (+0.61eV).

The metal oxidation state in all obtained Ta$_6$ cluster $H_2Ta_6O_{13}Cl_5^-$, $H_2Ta_6O_{14}Cl_3^-$, $H_2Ta_6O_{15}Cl^-$ is same. For all clusters we obtained two stable structures a trigonal bipyramid with additional Ta atom connected with polar Ta by two oxygen bridges and octahedron. In all cases the structure which stems from octahedron is the most stable. The stable structures are presented in Figures 13-15 and their bond lengths are presented in Tables 6,7.

MALDI spectra of Ta$_7$ clusters resulted in four following chemical compositions with the same Ta oxidation state: $H_3Ta_7O_{18}^-$ $H_3Ta_7O_{17}Cl_2^-$ $H_3Ta_7O_{16}Cl_4^-$ and $H_4Ta_7O_{15}Cl_6^-$. For these structures we investigated pentagonal structure, octahedron structure with attached one Ta atom via oxygen bridge and trigonal bipyramid structure with attached Ta atom in axial or equatorial positions. Figures 16-19 show stable theoretical structures. Pentagonal structure is stable for $H_3Ta_7O_{17}Cl_2^-$ (See Figure 17 d) only and among other stable structures for this formula it is less energetically stable. In first three cases the most energetically stable are clusters with attached two Ta atom to equatorial Ta atoms ( see Figures 16 a), 17 a) and 18 a) ).The second in energy stability series for $H_3Ta_7O_{18}^-$ is bipyramid with attached Ta atoms in axial positions (Figure 16 b) and for $H_3Ta_7O_{17}Cl_2^-$ and $H_3Ta_7O_{16}Cl_4^-$, is octahedral structure with attached Ta. Since Ta oxidation state in these clusters is less then maximal possible value 5, in the most stable forms there is metal-metal bonds, which are shown in Figure 20. Magnitudes of bond lengths are presented in Tables 8, 9, 10. In the case of Ta$_8$ only one composition $H_4Ta_8O_{18}Cl_5^-$ was obtained in MALDI spectra. For this structure we obtained cubic Ta$_8$ structure connected by oxygen bridges, with O, Cl atoms and OH group bonded with Ta atoms (see Figure 21). Magnitudes of bond lengths are presented in Table 11. Metal-metal bond is observed in this compound.

Table 8. Interatomic distances d(Å) in distorted pentagon of Ta$_7$ clusters.

| Figure | Ta$_{eq}$-Ta$_{eq}$ | Ta$_{eq}$-Ta$_{ax}$ | Ta-Cl | Ta-O$_{bridge}$ | Ta-OH |
|---|---|---|---|---|---|
| Fig 16 a) | 3.35(4) | 3.40(2) | | 1.96(2) | 1.889(5) |
| Fig 17 a) | 3.34(3) | 3.46(3) | 2.395(1) | 1.956(1) | 1.887(5) |
| Fig 18 a) | 3.16(5) | 3.34(1) | 2.345(3) | 1.984(8) | 1.881(6) |

Table 9. Interatomic distances d(Å) in trigonal bipyramid of Ta$_7$ clusters.

| Figure | Ta$_{eq}$-Ta$_{eq}$ | Ta$_{eq}$-Ta$_{ax}$ | Ta-Cl | Ta-O$_{bridge}$ | Ta-OH |
|---|---|---|---|---|---|
| Fig 16 b) | 3.268(2) | 3.526(3) | | 1.96(1) | 1.877(1) |
| Fig 17 c) | 3.270(3) | 3.526(1) | 2.364(1) | 1.96(2) | 1.875(1) |
| Fig 18 c) | 3.271(8) | 3.526(1) | 2.361(1) | 1.96(1) | 1.874 |

Table 10. Interatomic distances d(Å) in Ta₇ clusters based on octahedrons.

| Figure | Ta$_{eq}$-Ta$_{eq}$ | Ta$_{eq}$-Ta$_{ax}$ | Ta-Cl | Ta-O$_{bridge}$ | Ta-OH |
|---|---|---|---|---|---|
| Fig 16 c) | 3.46(3) | 3.46(6) |  | 1.96(1) | 1.870(1) |
| Fig 17 b) | 3.46(4) | 3.45(2) | 2.382(1) | 1.96(2) | 1.868(3) |
| Fig 18 b) | 3.491(1) | 3.41(4) | 2.328(3) | 1.95(2) | 1.866(1) |

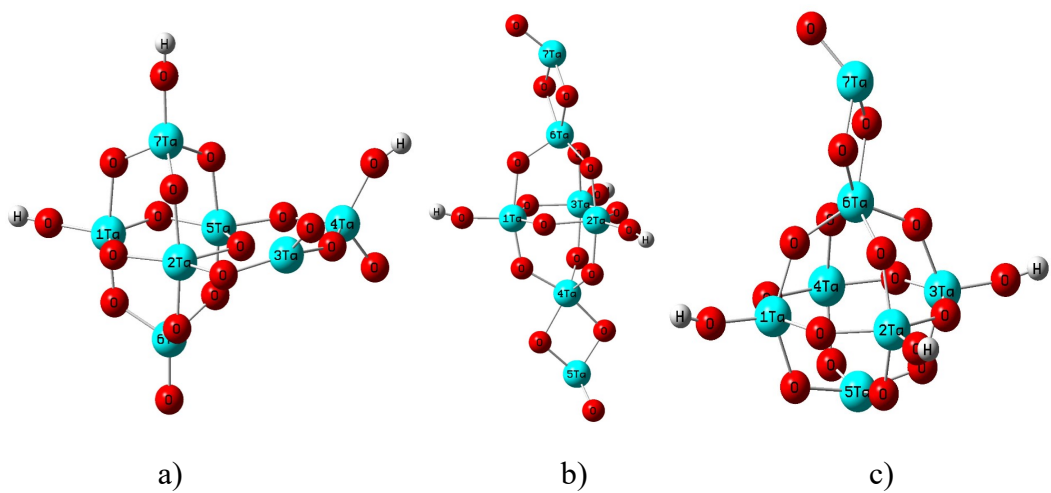

a)          b)          c)

Figure 16. Stable structures of $H_3Ta_7O_{18}^-$ and their relative total energies (in parenthesis) a) bipyramid structure with two additional Ta atoms laying in one plane (0.00 eV) b) bipyramid structure (+3.43eV) c) octahedron structure (+3.49eV).

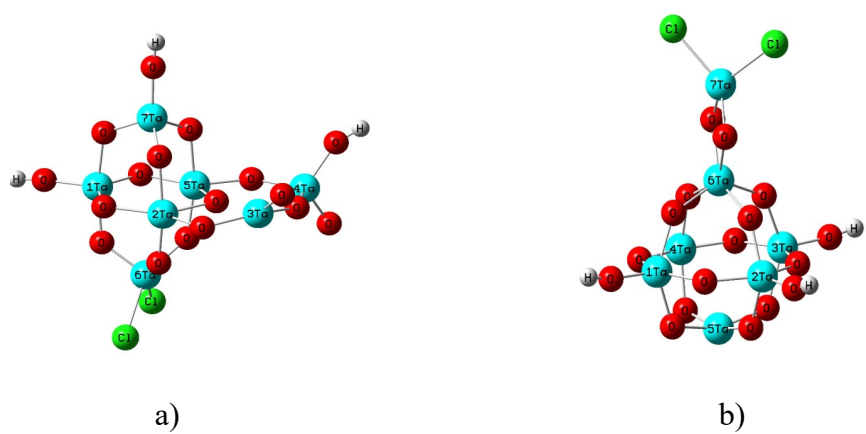

a)          b)

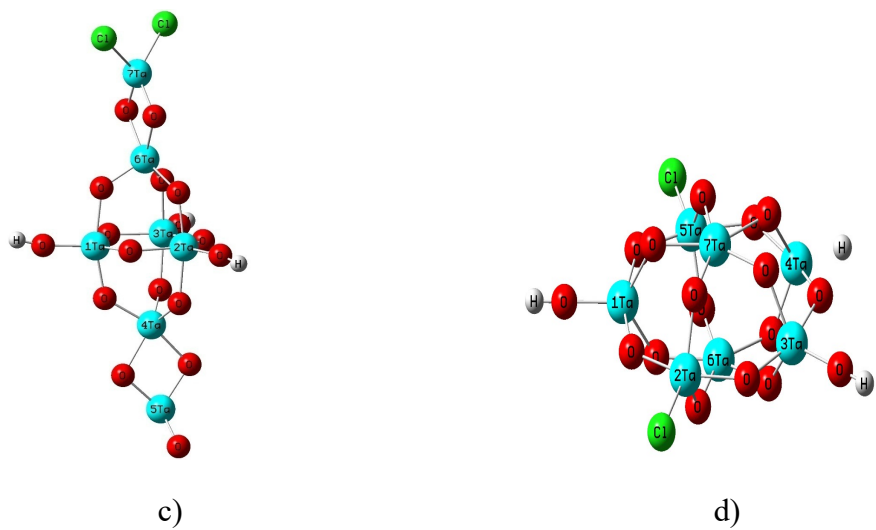

c)             d)

Figure 17. Stable structures of $H_3Ta_7O_{17}Cl_2^-$ and their relative total energies (in parenthesis) a) bipyramid structure with two additional Ta atoms laying in one plane (0.00 eV) b) octahedron structure (+2.34eV) c) bipyramid structure (+2.37eV) d) pentagon structure (+4.87eV).

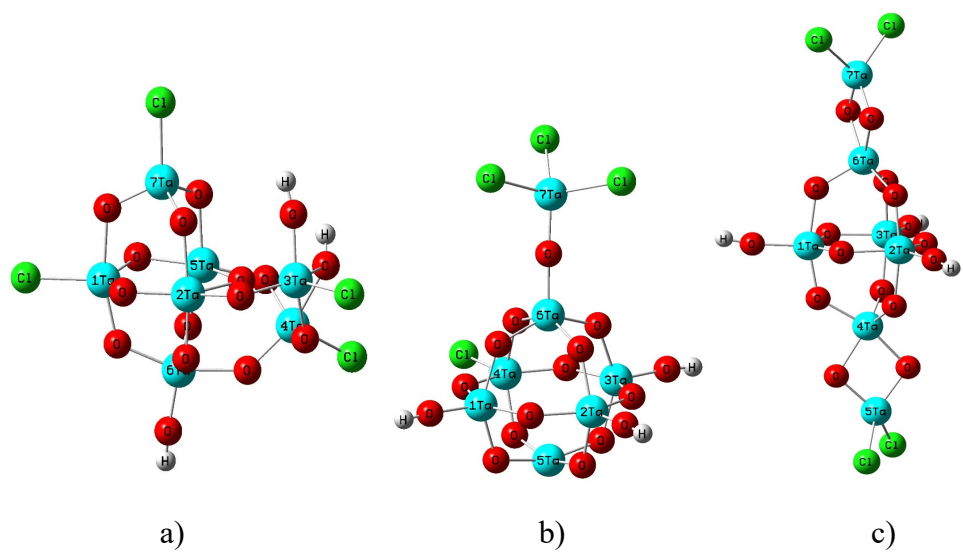

a)             b)             c)

Figure 18. Stable structures of $H_3Ta_7O_{16}Cl_4^-$ and their relative total energies (in parenthesis) a) bipyramid structure with two additional Ta atoms laying in one plane (0.00 eV) b) octahedron structure (+0.73eV) c) bipyramid structure (+1.85eV).

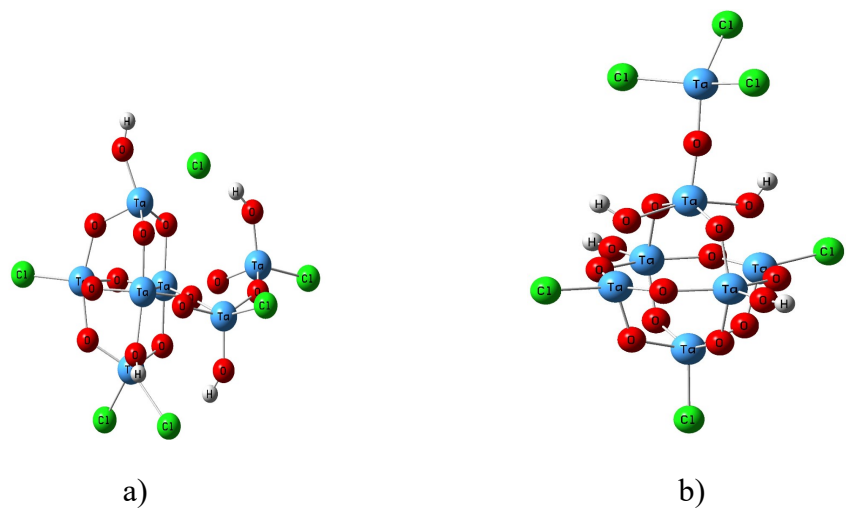

Figure 19. Stable structures of $H_4Ta_7O_{15}Cl_6^-$ and their relative total energies (in parenthesis) a) bipyramid structure with two additional Ta atoms laying in one plane (0.00 eV) b) octahedron structure (+0.87eV)

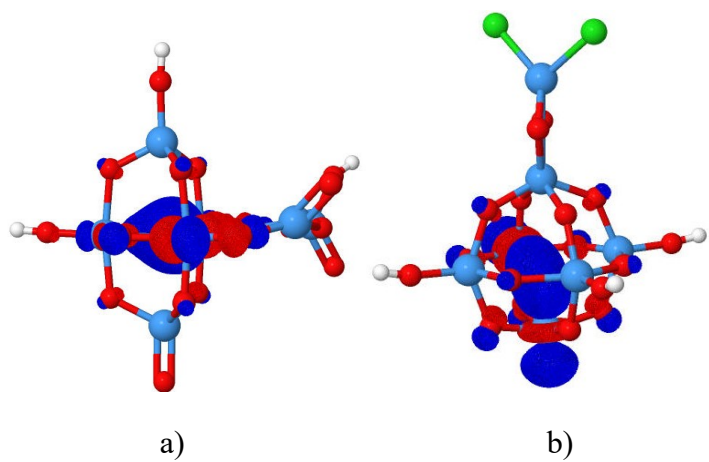

Figure 20. Orbitals representing metal-metal bonds in some $Ta_7$ structures a) $H_3Ta_7O_{18}^-$ bipyramid structure with two additional Ta atoms laying in one plane b) octahedron structure of $H_3Ta_7O_{17}Cl_2^-$

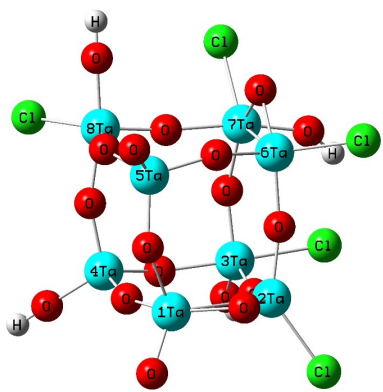

Figure 21. Cubic structure of $H_4Ta_8O_{18}Cl_5^-$

Table 11. Interatomic distances d(Å) in $Ta_8$ cluster.

| Figure | $Ta_{eq}$-$Ta_{eq}$ | $Ta_{eq}$-$Ta_{ax}$ | Ta-Cl | Ta-$O_{bridge}$ | Ta-OH |
|---|---|---|---|---|---|
| Fig 21 | 3.54(4) | 3.46(6) | 2.403(3) | 1.93(2) | 1.874(7) |

**Conclusion**

The MALDI time of flight and quantum chemical investigations of HTaOCl clusters in high oxidation state revealed the general features of their spatial and electronic structure. In the case of $Ta_3$ cluster triangular structure is more energetically stable than linear ones show metal-metal bonds. In the cases of $Ta_4$, and $Ta_5$ a clusters the most stable structures are distorted tetrahedron, trigonal bipyramid. Note, that the bipyramid structure was obtained to be the most stable with significant metal-metal bonds for $Mo_5O_8$ also.[9] Among of $Ta_6$ structures, having the same oxidation state the relative stability depends on the chemical constitution. Octahedral structure is the most stable for $H_2Ta_6O_{13}Cl_5^-$, $H_2Ta_6O_{14}Cl_3^-$, and the most stable $H_2Ta_6O_{15}Cl^-$ cluster resembles trigonal bipyramid with apical attachment of Ta via oxygen bridges. In the case of $Ta_7$ clusters, the most symmetrical stable structure of pentagonal bipyramid was obtained for $H_3Ta_7O_{17}Cl_2^-$ only, but its total energy is 4.87 eV higher than that of the most stable structure of trigonal bipyramid with attachment of two Ta atoms in equatorial position, which is also the most stable for all others except $H_3Ta_7O_{17}Cl_2^-$. For the last one the most stable is attachment of Ta atoms via oxygen bridges to apical Ta atoms of trigonal bipyramid. For $Ta_8$ only one cluster was registered in MALDI spectra, for which distorted cubic structure is stable. The core $Ta_8$ cube with 12 oxygen atoms in bridge positions resembles the $Mo_8O_{12}$ structure of $O_h$ symmetry. The distances between Ta atoms, connected by oxygen bridges are quite different, but if the Ta-Ta distance is less than 3.0 (Å) metal-metal bond is significant.

**Acknowledgements**

E.G.Il'in, A.S.Parshakov and V.G.Yarzhemsky acknowledge the support by IGIC RAS state assignment. E.A.Teplyakov acknowledges the support of numerical simulations of this work by the Russian Science Foundation (grant RNF 18-12-00438).